\documentclass[manuscript]{aastex}
\usepackage{color}
\usepackage{subfigure}
\usepackage{epsfig}
\usepackage{lscape}
\usepackage{soul}
\usepackage[normalem]{ulem}

\shorttitle{Carbon and Oxygen Abundances in the Hot Jupiter Exoplanet Host Star}
\shortauthors{Teske et al.}

\begin{document}

\newcommand{\txw}{\textwidth}
\title{Carbon and Oxygen Abundances in the Hot Jupiter Exoplanet Host Star XO-2N and its Binary Companion\altaffilmark{*}}

\altaffiltext{*}{Based on data collected at Subaru Telescope, which is operated by the National Astronomical Observatory of Japan.}

\author{Johanna K. Teske\altaffilmark{1}, Simon C. Schuler\altaffilmark{2}, Katia Cunha\altaffilmark{1, 3}, Verne V. Smith\altaffilmark{4}, Caitlin A. Griffith\altaffilmark{5}}

\altaffiltext{1}{Steward Observatory, University of Arizona, Tucson, AZ, 85721, USA; email: jteske@as.arizona.edu}
\altaffiltext{2}{University of Tampa, 401 W. Kennedy Blvd., Tampa, FL 33606, USA}
\altaffiltext{3}{Observat\'orio Nacional, Rua General Jos\'e Cristino, 77, 20921-400, S\~ao Crist\'ov\~ao, Rio de Janeiro, RJ, Brazil}
\altaffiltext{4}{National Optical Astronomy Observatory, 950 North Cherry Avenue, Tucson, AZ 85719, USA}
\altaffiltext{5}{Lunar and Planetary Laboratory, University of Arizona, Tucson, AZ, 85721, USA}

\clearpage

\begin{abstract}
With the aim of connecting the compositions of stars and planets, we
present the abundances of carbon and oxygen, as well as iron and
nickel, for the transiting exoplanet host star XO-2N and its
wide-separation binary companion XO-2S.
Stellar parameters are derived 
from high-resolution, high-signal-to-noise spectra, and the two stars
are found to be similar in their T$_{\rm{eff}}$, log $g$, iron
([Fe/H]), nickel ([Ni/H]) abundances. 
Their carbon ([C/H]) and oxygen ([O/H]) abundances also overlap within errors, although 
XO-2N may be slightly 
more C-rich and O-rich than
XO-2S. The C/O ratios of both stars ($\sim$0.60$\pm$0.20) may also be somewhat larger than solar 
(C/O$\sim$0.50). 
%
The XO-2 system 
has a transiting hot Jupiter 
orbiting one binary
component but not the other, allowing us to probe the potential
effects planet formation might have on the host star
composition. Additionally, 
with multiple observations of its
  atmosphere the transiting exoplanet XO-2b lends itself to
  compositional analysis, which can be compared to the natal chemical
environment established by our binary star elemental abundances. 
This work sets the stage for determining how similar/different
exoplanet and host star compositions are, and the implications for
planet formation, by discussing the 
C/O ratio measurements in the unique environment of a visual binary system 
with one star hosting a transiting hot Jupiter. 

\end{abstract}


\section{Introduction}
Although theory and observations indicate that host-star composition
affects planetary evolution, the physical processes responsible are
not well understood. Until recently, investigations of the chemical connection between stars and planets were limited to measurements of 
the host star abundances. 
One of the most prominent findings is that the
(solar-type) host stars of large, closely orbiting (hot Jupiter)
exoplanets are more metal-rich than (solar-type) stars without
detected gas giant exoplanets (e.g., Gonzalez et al.\,1998; Santos et
al.\,2004; Fischer \& Valenti 2005). However, the host star
metallicity trend is weaker for Neptune-sized planets (e.g., Ghezzi et al.\,2010) and has been found to \textit{not} hold for terrestrial-sized planets (Buchhave et al.\,2012), whose host stars show a wide range of metallicities. 

With the discovery of transiting exoplanets, planetary atmospheres
themselves can be observed, and their compositions determined. Indeed,
the \textit{Hubble Space Telescope} and \textit{Spitzer Space
  Telescope} have been used to detect the most abundant molecules (H$_2$O, CO, CH$_4$, CO$_2$) in the atmospheres of several of the brightest transiting planets (e.g., Tinetti et al.\,2007; Swain et al.\,2008; D\`{e}sert et al.\,2009).
Measurements of both stellar and exoplanetary atmospheres combined 
provide valuable insight into planet formation processes. 
The ratio of carbon to oxygen is important to
  interpreting hot Jupiter exoplanet spectra because they are dominated
  by the main carbon- and oxygen-containing molecules and 
particularly sensitive to 
different chemistry induced by different C/O ratios. 
In thermochemical equilibrium, at the temperatures and pressures characteristic of gas giant
atmospheres, a high C/O ratio causes differences in the partitioning of
C and O among H$_2$O, CO, CH$_4$, CO$_2$ compared to that expected in
solar abundance atmospheres (C/O$_{\odot}$=0.55$\pm$0.10; Asplund et
al.\,2009; Caffau et al.\,2011) (Kuchner \& Seager 2005; Kopparapu et al.\,2012; Madhusudhan
2012). This in turn affects the composition and thermal structure, and
therefore spectral signatures, of exoplanet atmospheres. Currently a
number of teams are working towards establishing the C/O ratios of
transiting exoplanets (e.g., Madhusudhan 2012). Observations of the
host stars are needed to interpret the exoplanet observations in the context of
the elemental composition of each star-planet system.

The C/O ratio of an exoplanet can also give clues as to where in the protoplanetary disk it formed (Stevenson \& Lunine 1988; \"Oberg et al.\,2011). Observations indicate that disks are inhomogenous in physical structure and composition (e.g., Bergin 2011), 
but that in particular carbon and oxygen in a planet's atmosphere could 
be indicative of its starting orbital position and evolution (\"Oberg et al.\,2011). Some studies suggest that a planet may also affect the elemental composition of the host star (e.g., Mel\'endez et al.\,2009; Ram\'{\i}rez et al.\,2009). It is important to be able to isolate these two potential effects -- the effect of the planet on the star's elemental abundances, and the starting (or unperturbed) abundance of the system from which one can study a planet's origin and evolution. Studies of binary star systems, as conducted here, provide a method for decoupling these two potential effects.

Reported in this paper 
are the C/O\footnote{The C/O ratio -- the ratio of carbon atom
   to oxygen atoms -- is calculated in stellar abundance analysis as C/O=
  N$_{\rm{C}}$/N$_{\rm{O}}$=10$^{\rm{logN(C)}}/10^{\rm{logN(O)}}$ where
  log(N$_{\rm{X}}$)=log$_{10}$(N$_{\rm{X}}$/N$_{\rm{H}}$)+12.} ratios of a transiting exoplanet host star and its binary companion. We describe the XO-2 binary system, which consists of the hot
Jupiter host XO-2N, and its 
companion XO-2S, located $\sim$4600 AU
away 
and 
not known to host a Hot Jupiter-type planet (Burke et al.\,2007). 
The hot Jupiter XO-2b has an M$\times$sin$i$ of
0.62$\pm$0.02 M$_{\rm{J}}$\footnote{$i=88.7\pm1.3\,^{\circ}$, meaning sin\,$i$ is nearly unity} (Narita et al.\,2011), a radius of
0.97$\pm$0.03 R$_{\rm{J}}$ (Burke et al.\,2007), and orbits at
$\sim$0.04 AU from XO-2N. The exoplanet XO-2b is one of the best
characterized bodies outside the solar system, studied extensively with HST and \textit{Spitzer}
(e.g., Machalek et al.\,2009; Crouzet et al.\,2012). We perform a
stellar abundance analysis of both binary components to
investigate the potential chemical effects of exoplanet formation.

\section{Observations and Data Reduction}

Observations of XO-2N and XO-2S were conducted during two half-nights, February 10 and 11 2012 (UT), with the 8.2 m Subaru Telescope using the High Dispersion Spectrograph (HDS; Noguchi et al.\,2002).
Spectra of the Sun (as reflected moonlight) were taken the first night, and spectra of a telluric standard (HR 6618) were taken on both nights. A 0.''6 slit width 
was used, providing a resolution of
$R= \frac{\lambda}{\Delta\lambda} = 60,000$, with two-pixel binning in
the cross-dispersion direction and no binning in the dispersion
direction. Across the two detectors, wavelength coverage of the
spectra is $\sim$4450\,{\AA}-5660\,{\AA} and
$\sim$5860\,{\AA}-7100\,{\AA}. The signal-to-noise (S/N) ratios in the combined 
frames ranged
from $\sim$170-230. The 
raw data were reduced using standard techniques within the IRAF\footnote{IRAF is distributed by the National Optical Astronomy Observatory, which is operated by the Association of Universities for Research in Astronomy, Inc., under cooperative agreement with the National Science Foundation.} software package. 

\section{Abundance Analysis and Results}
Stellar parameters (T$_{\rm{eff}}$, log $g$,
microturbulence [$\xi$]) and elemental abundance ratios were derived
following the procedures in Schuler et al.\,(2011a) and Cunha et
al.\,(1998). We used the spectra themselves to determine the
  parameters by forcing zero correlation between [Fe I/H] and lower
  excitation potential ($\chi$), and between [Fe I/H] and reduced
  equivalent width [log(EW/$\lambda$)], as well as ensuring that the
  [Fe/H]\footnote{We use the standard bracket notation to indicate abundances relative to solar, e.g., [X/H]=log(N$_{\rm{X}}$) - log (N$_{\rm{X}}$)$_{\rm{solar}}$} abundances derived from Fe I and Fe II lines were equal to
  within two significant figures. The abundances were determined using an updated version
of the local thermodynamic equilibrium (LTE) spectral analysis code
MOOG (Sneden 1973), with model atmospheres interpolated from the
Kurucz ATLAS9 grids\footnote{See
  http://kurucz.harvard.edu/grids.html}. All abundances were
normalized to solar values on a line-by-line basis. Abundances of Fe, Ni, and C
were derived directly from equivalent width (EW) measurements of
spectral lines in each target (with the ``abfind'' driver in MOOG).
The EW measurements were performed with either the one-dimensional
spectrum analysis package SPECTRE (Fitzpatrick \& Sneden 1987) or the
`splot' task in IRAF. 

Specifically, Fe lines were chosen from the Schuler et al.\,(2011a) line list. We measured 49 Fe I lines in XO-2N and XO-2S, and 8 and 10 Fe II lines in
XO-2N and XO-2S, respectively. 
Lower excitation potentials 
and transition probabilities (log $gf$) were taken from the Vienna
Atomic Line Database (VALD; Kupka et al. 1999) for Fe, C, and Ni. 

\begin{figure}[ht!]
\figurenum{1}
\centering
\includegraphics[width=.75\textwidth, angle =-90]{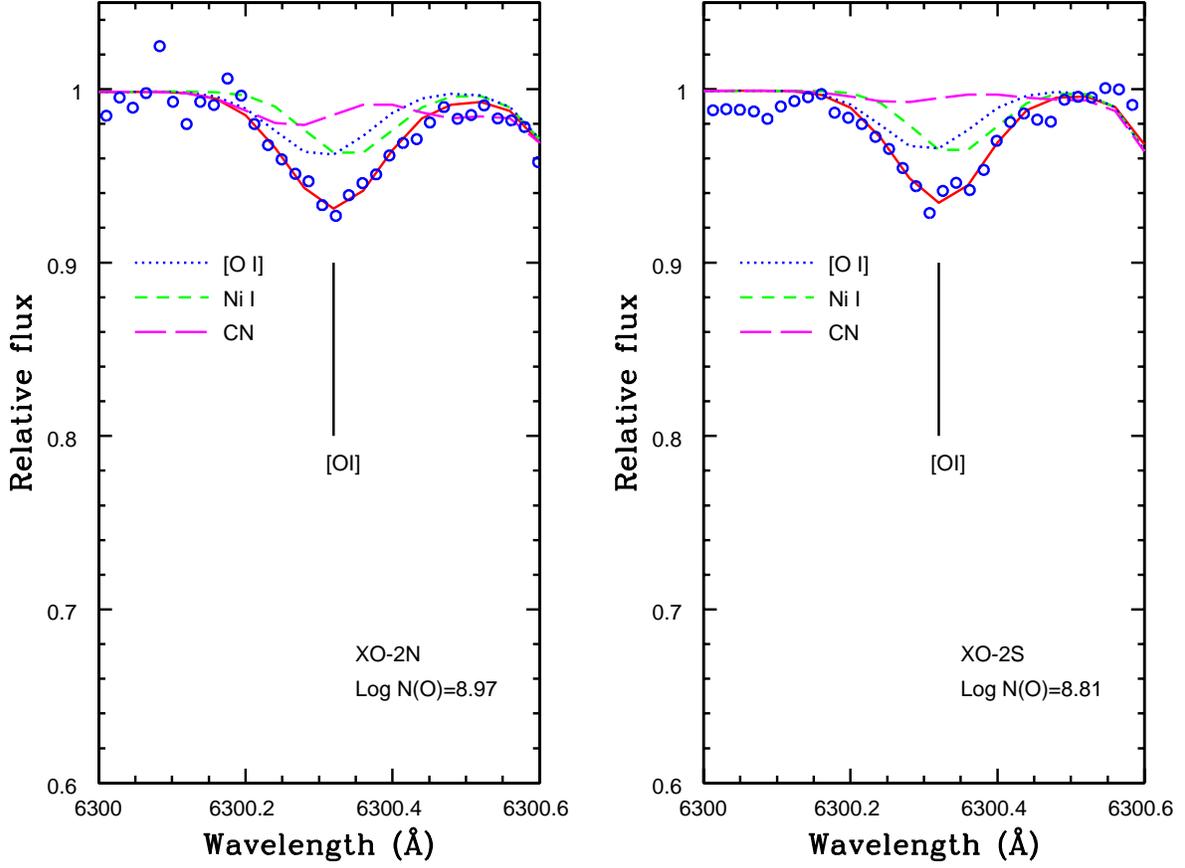}
\caption{Shown are the spectrum synthesis fits to the forbidden [OI]
  line (6300.3 {\AA}) for XO-2N (left) and XO-2S (right). The data are
shown as blue open circles. The full synthesis fit is shown with a solid
  red line, with components shown with
blue dotted ([OI]), green short-dashed (Ni I), and magenta long-dashed
(CN) lines. Recall log(N$_{\rm{X}}$)=log$_{10}$(N$_{\rm{X}}$/N$_{\rm{H}}$)+12.}
\label{synth_fig}
\end{figure}

Carbon abundances for XO-2N and XO-2S were derived from two C I lines at 5052\,{\AA} and 5380\,{\AA}, 
which have been shown to provide reliable abundances in solar-type stars (Takeda \& Honda 2005). 
Oxygen abundances were derived from the forbidden [O I] line at
$\lambda$ = 6300.3\,{\AA}, which is well-described by LTE (e.g. Takeda
2003); we used the Allende Prieto et al.\,(2001) log $gf$ value of [O
I] line. 
Our analysis of the forbidden oxygen line used the spectrum synthesis
method (with the ``synth'' driver in MOOG; see Figure\,\ref{synth_fig}) to account for its
blending with a Ni I line and a CN line. 
The free parameters of the synthesis
fit were the continuum normalization, wavelength shift
(left/right), line broadening, and oxygen abundance; we used 
our measured Ni and C abundances for each star, and N scaled from
solar based on
the measured [Fe/H] of each star (e.g., Cunha et al.\,1998). The Ni I line is composed of two isotopic components; the weighted log $gf$ values of the two components from 
Bensby et al.\,(2004) were used here. We note that the [O I] line in XO-2N (with planet) required more broadening to fit with the synthesis method, suggesting it has a larger \textit{v sini} than the planet-less XO-2S.

Uncertanties in T$_{\rm{eff}}$ and
$\xi$ were calculated by forcing 1$\sigma$ correlations in the
relations between [Fe I/H] and $\chi$ and between [Fe I/H] and reduced
equivalent width [log(EW/$\lambda$)], respectively. The change in T$_{\rm{eff}}$ or $\xi$
required to cause a correlation coefficient $r$ significant at the
1$\sigma$ level was adopted as the uncertainty in these
parameters. The uncertainty in log $g$ was calculated differently,
through an iterative process described in detail in Baubar \&
King\,(2010) and outlined in Schuler et al.\,(2011a). 

There are two components to the uncertainties in the derived elemental
abundances -- one from stellar parameter errors 
and one from the dispersion 
in the abundances derived from different elemental
absorption lines. 
To determine the uncertainty due to the stellar parameters, the
sensitivity of the abundance to each parameter was calculated for
changes of $\pm150$ K in T$_{\rm{eff}}$, $\pm0.25$ dex in log $g$, and
$\pm0.30$km s$^{-1}$ in $\xi$. The final uncertainty due to each
parameter is then the product of this sensitivity and the
corresponding parameter uncertainty (as described above). The second
uncertainty component 
is parameterized with the uncertainty in the mean, $\sigma_{\mu}$\footnote{$\sigma_{\mu}=\sigma/\sqrt{N-1}$,
where $\sigma$ is the standard deviation of the derived abundances and
$N$ is the number of lines used to derive the abundance.}, for the
abundances derived from the averaging of multiple lines. Then the
total 
uncertainty for each abundance ($\sigma_{\rm{tot}}$) is the quadratic
sum of the (3) individual parameter uncertainties and $\sigma_{\mu}$.

The equivalent width measurements from our analysis are shown in
Table \ref{tab:lines}, along with the wavelength, $\chi$, log $gf$,
EWs, and line-by-line abundances for each element for the Sun, XO-2N,
and XO-2S. The final derived stellar parameters and
their 1$\sigma$ uncertainties, as well as the derived [Fe/H], [C/H], [Ni/H], [O/H],
and C/O ratio values and their 1$\sigma$ uncertanties, are shown in
Table \ref{tab:stellar_params}. The C/O ratio errors are the errors of
[C/H] and [O/H] combined in quadrature. 
We find [C/H]$=$ +0.26$\pm$0.11 in XO-2S versus +0.42$\pm$0.12 in
XO-2N, and [O/H]$=$ +0.18$\pm$0.15 in XO-2S versus +0.34$\pm$0.16 in
XO-2N.  

In addition to our results, Table \ref{tab:stellar_params} lists stellar parameters and [Fe/H] of Ammler-von Eiff et al.\,(2009) and Torres et al.\,(2012), two studies comparable to this one in their analysis methods and sample. Also listed are results 
from the exoplanet discovery paper, Burke et al.\,(2007). Instead of the MOOG+Kurucz models as implemented here, Burke et al.\,(2007) used the Spectroscopy Made Easy (SME) code (Valenti \& Piskunov 1996), which has been demonstrated to be biased by correlations bewteen T$_{\rm{eff}}$, [Fe/H], and log $g$ 
as compared to a MOOG analysis (Torres et al.\,2012). 
Their results are shown because Ammler-von Eiff et al.\,(2009) and Torres et al.\,(2012) do not include XO-2S. 
Our results for XO-2S differ slightly from those of Burke et al.\,2007 (based on SME analysis), but 
our results for XO-2N are more consistent with those of Burke et al.\,2007
, and the same within errors as Ammler-von Eiff et al.\,(2009) and Torres et al.\,(2012), who found similarly large T$_{\rm{eff}}$ error for XO-2N. 

\section{Discussion}

XO-2 stands out among transiting exoplanet systems because its host
star, XO-2N, is in a wide binary. 
XO-2S, without a large, close-in planet, can be
studied to determine the composition of the unperturbed environment in
which these stars and planet(s) formed.
We find that XO-2S
and XO-2N are similar in their physical properties (with
XO-2S being slightly more massive based on T$_{\rm{eff}}$ and log
$g$), as well as [Fe/H] and [Ni/H]. The errors in [C/H] and [O/H] (both relative to solar)
allow for larger differences in the stars' respective carbon and
oxygen abundances (see Table \ref{tab:stellar_params}), though within
errors they also overlap. 
Schuler et al.\,(2011b)
conducted a similar analysis (though did not measure O) 
for another
roughly-equal-mass binary with one component hosting a giant planet, 16
Cyg A and B, and found the two stars to be chemically homogeneous
(aside from Li and B, attributed to different internal mixing
efficiencies). Our results for the XO-2 transiting planet system also
indicate that the stars are chemically alike -- 
here 
we additionally determine the binary stars' C/O ratios and find both to have C/O$\sim$0.60.
It is currently unclear whether and how planets affect the composition
of the host star 
(e.g., Mel\'endez et al.\,2009; Ram\'{\i}rez et al.\,2009;
Chambers 2010; Gonz\'alez Hern\'andez et al.\,2010; Schuler et
al.\,2011a \& 2011b). Several studies posit that stars with smaller planets are 
depleted in rock-forming (refractory, e.g. Mg, Si, Ni, Al)
elements relative to volatile elements (e.g., C, N, O) due to
rock-forming material being ``locked up'' in the terrestrial planets
(e.g., Mel\'endez et al.\,2009; Ram\'{\i}rez et al.\,2009). The Sun has been shown 
to be deficient (by $\sim$20\%) in refractory elements that have $T_c\gtrsim$ 900 K relative to
volatile elements compared to similar stars without detected
planets (e.g., Mel\'endez et al.\,2009; Ram\'{\i}rez et
al.\,2009). However, the details of how an individual star's
atmosphere is affected by the local or global composition of the disk during
its evolution are uncertain. Also, we do not know whether some or even most
stars hosting detected hot Jupiters actually also have small planets that
might cause such a signature. The significance of our detection in
both XO-2N and XO-2S of enhanced [Fe/H] and [Ni/H], the only two
refractory abundances measured here, will be better understood when
compared to a larger number of refractory elements measured in this
system and the abundance trends with $T_{c}$ expected based on
galactic chemical evolution. 

No previous study
of which we are aware has uniformly derived [C/H], [O/H], and
C/O 
values for this binary system. 
Several studies have examined C/O ratios in non-transiting
exoplanet host stars versus 
stars without known exoplanets. [It should be noted that any star
designated as a ``non-host'' has the potential to 
harbor a smaller (
undetected) planet; indeed, it may be the case that most stars have
one or more small planets (e.g., Cassan et al.\,2012).] Delgado Mena
et al.\,(2010) measured carbon (using high excitation C I lines) and
oxygen 
(using the forbidden [OI] 6300\,{\AA} line) in 100 giant planet host
stars from the HARPS planet-search sample, along with 270 non-host
stars. They found averaged host-star values of [C/H]=$+$0.10$\pm$0.16,
[O/H]=$+$0.05$\pm$0.17, and C/O=0.76$\pm$0.20, with corresponding
``single'' star (no known planets) averages of [C/H]=$-$0.06$\pm$0.18,
[O/H]=$-$0.08$\pm$0.17, and C/O=0.71$\pm$0.18. Similar averages,
overlapping within 1$\sigma$ errors, were compiled by Bond et
al.\,(2010) and measured by Petigura \& Marcy\,(2011). These studies
show 
no significant difference in [C/H], [O/H], or C/O between stars
with/without detected (non-transiting) exoplanets. 

Fortney\,(2012) suggested the C/O ratios of both host- and non-host
stars in these studies were overestimated due to errors in the derived
C/O ratios and the observed frequency of carbon dwarf
stars in large samples of low-mass stars. More recently, Nissen\,(2013)
  determined C/O for 33 host stars from the Delgado Mena et
  al.\,(2010) sample that had additional ESO 2.2m FEROS spectra
  covering the O I triplet at 7774\,{\AA} (unavailable in our data). 
Accounting for non-LTE effects on the [OI]
  triplet, Nissen\,(2013) found 
differences in
  derived [O/H] as compared to Delgado Mena et
  al.\,(2010), resulting in a 
tighter correlation between [Fe/H]
  and the C/O ratios derived by Nissen (C/O = 0.58$+$0.48[Fe/H] with
  an rms dispersion $\sigma$(C/O)=0.06). However, the averaged
  host-star values of Nissen,\,(2013) overlap those of Delgado Mena et
al.\,(2010) and the other studies listed above ([C/H]=$+$0.11$\pm$0.15,
[O/H]=$+$0.08$\pm$0.10, and C/O=0.63$\pm$0.12).

\begin{figure}[hbt]
\figurenum{2}
   \subfigure{\includegraphics[width=0.5\textwidth]{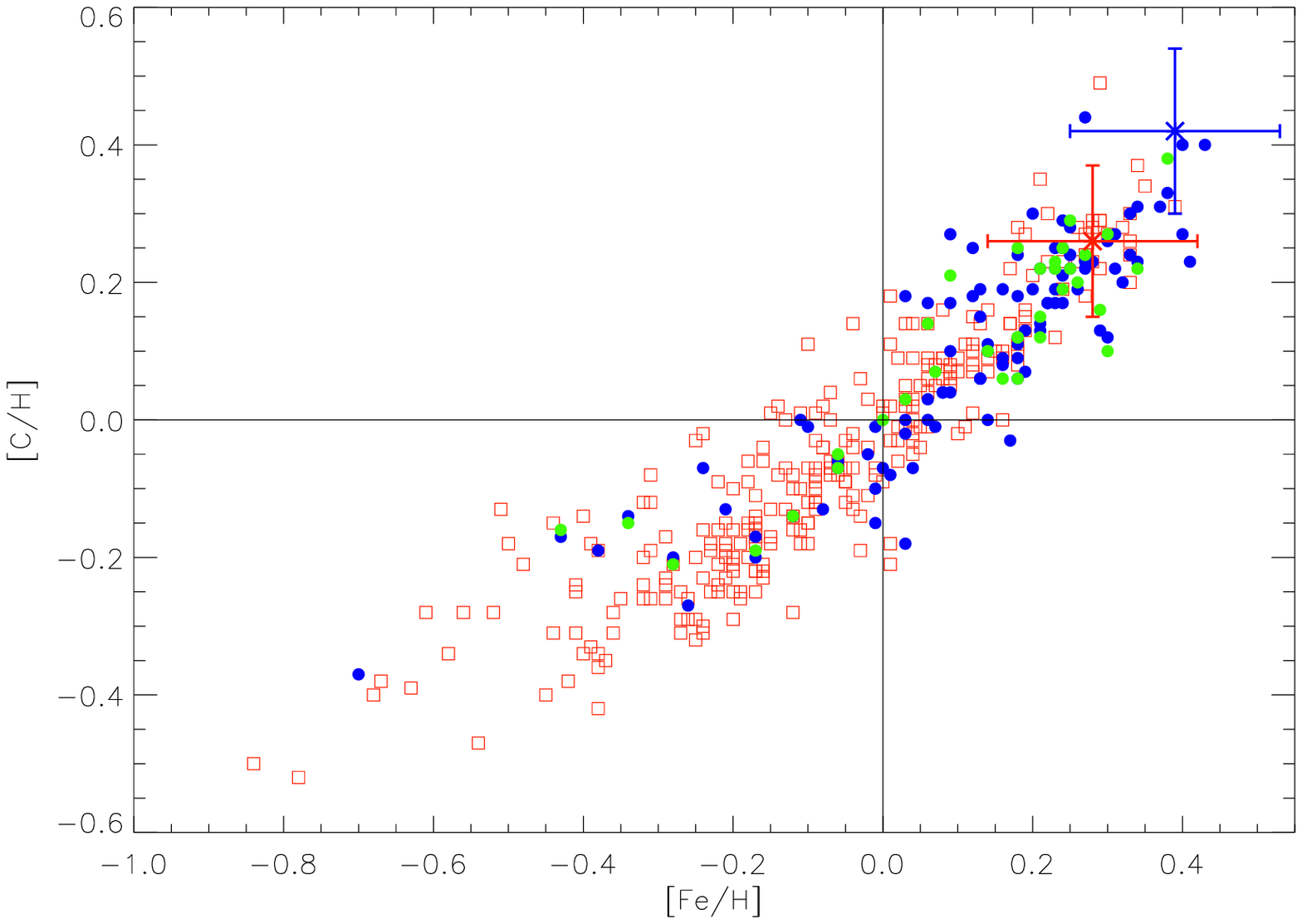}}
\quad
   \subfigure{\includegraphics[width=0.5\textwidth]{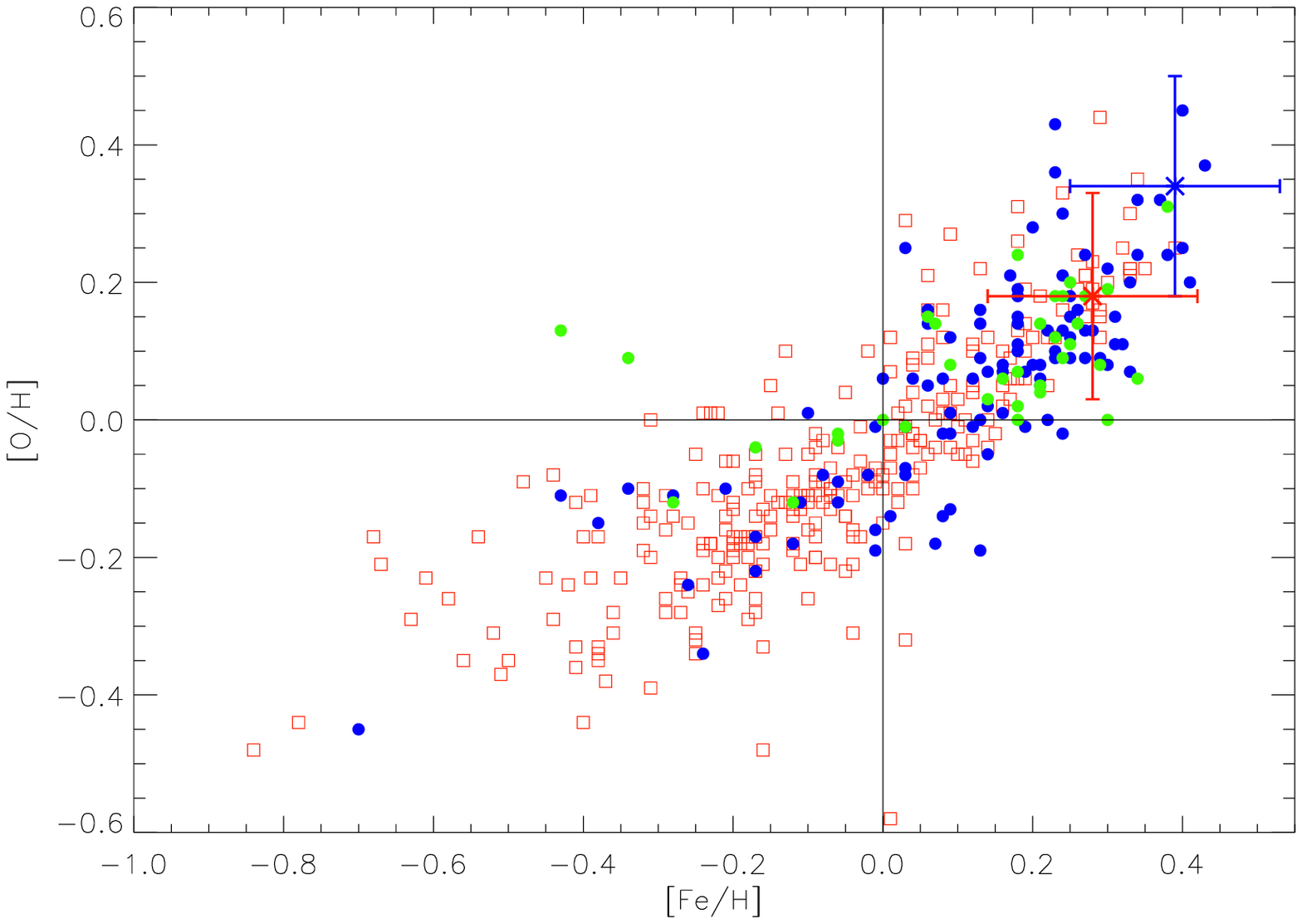}}
\quad
   \subfigure{\includegraphics[width=0.5\textwidth]{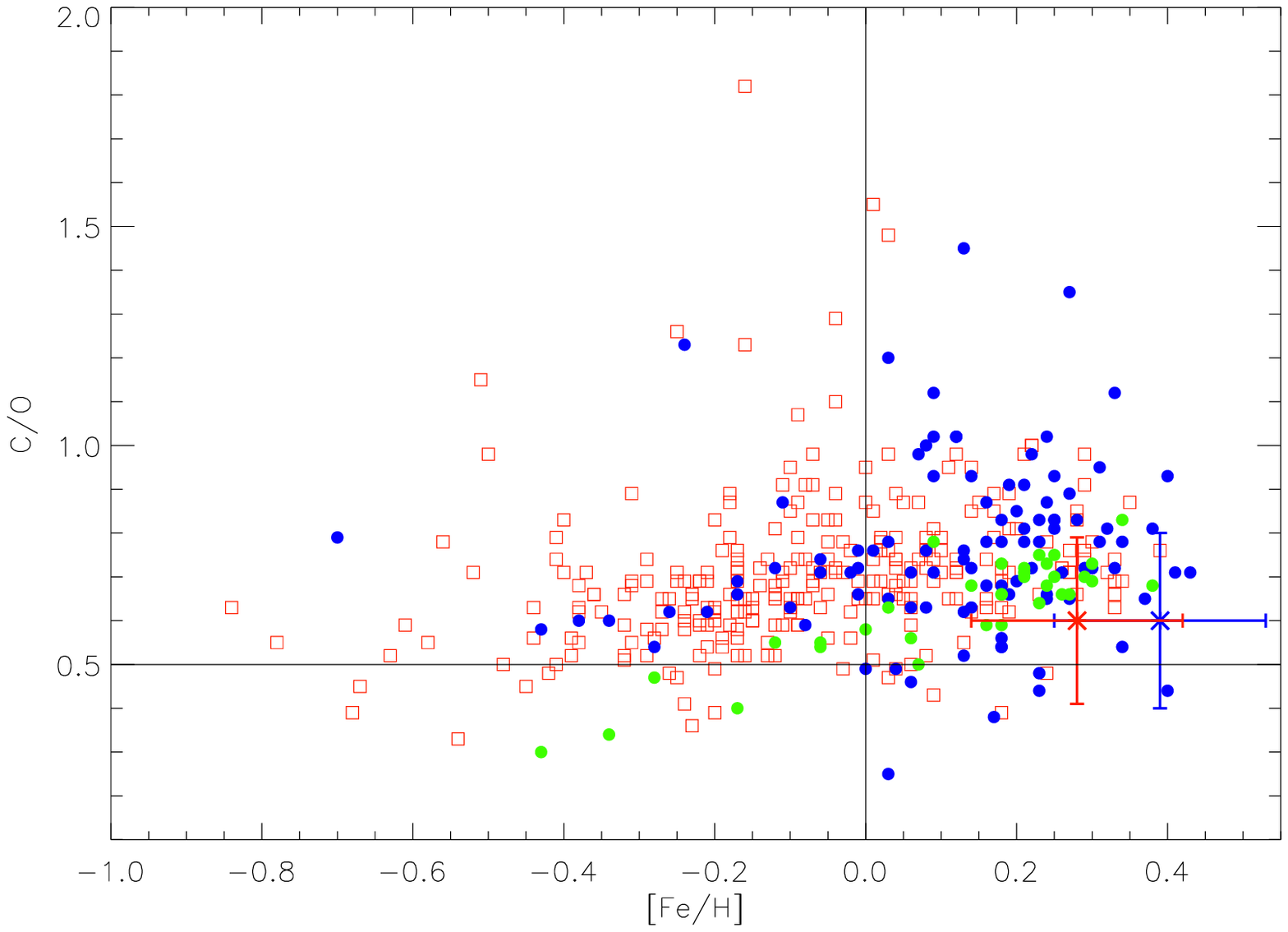}}
\caption{[C/H], [O/H], and C/O versus
  [Fe/H] from Delgado Mena et al.\,(2010) and Nissen\,(2013) [all
  Nissen\,(2013) hosts are in the Delgado Mena et al.\,(2010) host sample]. Non-host stars from Delgado Mena et al.\,(2010) are plotted with red open squares, while host stars from Delgado Mena et al.\,(2010)/Nissen\,(2013) are plotted with blue/green circles. Measurements of XO-2N (blue) and XO-2S (red) from this work are plotted as asterisks, with error bars included (see Table \ref{tab:stellar_params}). Black solid lines show the solar values.}
\label{fig2}
\end{figure}

The [C/H] values derived here for XO-2S (+0.26$\pm$0.11) and XO-2N
(+0.42$\pm$0.12) are 
both larger than
the averages 
above, and greater than
solar. 
These stars are also enhanced in [O/H] (see Table 2)
and have C/O ratios of 
0.60. 
This is the first measurement of the C/O ratio in a 
transiting exoplanet host star that is metal-rich. 
Since XO-2N and XO-2S are physically associated, the elevated [C/H] and [O/H] values in 
\textit{both} stars are strong evidence that their parent molecular cloud was
elevated in both carbon and oxygen. 
As shown in Figure~\ref{fig2}, XO-2S and XO-2N follow the broad
galactic chemical evolution trends in [C/H], [O/H], and C/O versus
[Fe/H] as evidenced by the large sample of Delgado Mena et
al.\,(2010) (and also Nissen 2013). 
The significance of the C/O ratio derived here is supported by the
careful analysis and the fact that we measured this ratio in two
separate stars within the same system.

The C/O ratio of a planet does not necessarily reflect the
protoplanetary-disk-averaged C/O ratio, and depends 
on where formation occurs, how much of the atmosphere is accreted from
gas versus solids, and how isolated the atmosphere is from the core
(\"Oberg et al.\,2011). 
Carbon-enhanced systems may actually have more solid mass in the inner disk than solar-composition environments, due to a wide
zone of C-bearing solids close to the star and a paucity of water ice
farther out in the disk (Bond et al.\,2010). 
If planets do form  more readily
in C-rich environments, this might help explain why there is a giant
planet around XO-2N ([C/H]=+0.42$\pm$0.12) and not XO-2S ([C/H]=+0.26$\pm$0.11). 
However, within uncertainties, 
the two stars' carbon and oxygen
abundances overlap. 
Furthermore, Delgado Mena et al.\,(2010) suggest that, based on a lack of trends between C/O ratios and planetary period, semi-major axis, and mass, any effects of an alternative mass distribution due to 
C-rich material is quickly erased. Thus the key to understanding why XO-2N has a planet and XO-2S does not may lie in the exoplanet composition.

\section{Conclusions}

We present an abundance analysis for the transiting exoplanet host
star XO-2N and its wide-separation binary companion XO-2S. The two
stars are found to be similar in their physical and chemical
properties, and both enhanced above solar in carbon and oxygen, with
C/O$\sim$0.6. Insight into why XO-2N hosts a transiting hot Jupiter,
and XO-2S does not, may be revealed by the atmospheric composition and
C/O ratio of the planet, which is currently being constrained with
observations recorded during the its primary/secondary eclipse (e.g.,
Machalek et al.\,2009; Crouzet et al.\,2012). While previous work
suggests that refractory element distributions may differ in stars
with/without planets, differences in volatile elements have not been
as thoroughly explored. Our results motivate further studies of
planet formation and evolution with a renewed focus on volatile
element distributions (particularly C and O) in both gas giant and
terrestrial planets. Additional measurements of binary host star
compositions will 
connect exoplanets and their stars and 
expand upon the giant planet-metallicity trend to investigate how host star chemical composition (especially C/O) influences planet formation and composition. 


\acknowledgements
The authors wish to recognize and acknowledge the very significant cultural role and reverence that the summit of Mauna Kea has always had within the indigenous Hawaiian community. We are most fortunate to have the opportunity to conduct observations from this mountain. This work would not have been possible without the efforts of the daytime and nighttime support staff at the Mauna Kea Observatory and Subaru Telescope. The work of J.T. and C.G. is suppored by NASA's Planetary Atmospheres Program. 
We also thank the anonymous referee for her/his helpful comments, and Jeremy King for several insightful discussions. 

{\it Facilities:} \facility{Subaru}


\begin{deluxetable}{lcccccccccc}
\tablecolumns{11}
\tablewidth{0pc}
\tabletypesize{\scriptsize}
\tablecaption{Lines Measured, Equivalent Widths, and Abudances \label{tab:lines}}
\tablehead{ \colhead{Ion} & \colhead{$\lambda$} & \colhead{$\chi$} & \colhead{log $gf$} & \colhead{EW$_{\odot}$} & \colhead{log$N_{\odot}$} & \multicolumn{2}{c}{XO-2S} & \colhead{ } & \multicolumn{2}{c}{XO-2N} \\ 
  \cline{7-8} \cline{10-11} \\
  \colhead{ } & \colhead{({\AA})} & \colhead{(eV)} & \colhead{(dex)} & \colhead{(m{\AA})} & \colhead{ } & \colhead{EW (m{\AA})} & \colhead{log $N$} & \colhead{ } & \colhead{EW ({\AA})} & \colhead{log $N$}}
\startdata
C I & 5052.17 & 7.68 & -1.304 & 27.1 & 8.31 & 36.4 & 8.61 & \colhead{ } & 33.0 & 8.77  \\ 
{ } & 5380.34 & 7.68 & -1.615 & 18.6 & 8.41 & 22.4 & 8.62 & \colhead{ } & 20.5 & 8.79  \\ 
O I & 6300.30 & 0.00 & -9.717 & 5.4 & 8.63 & 10.7 & 8.81 & \colhead{ } & 11.5 & 8.97  \\
\enddata
\tablecomments{This table is available in its entirety in a machine-readable form in the online journal. A portion is shown here for guidance regarding its form and content.}
\end{deluxetable}

\begin{deluxetable}{lccccccc}
\tabletypesize{\scriptsize}
\tablecolumns{8}
\tablewidth{0pc}
\tablecaption{Derived Stellar Parameters \label{tab:stellar_params}}
\tablehead{ 
  \colhead{Parameter$^{a}$} & \multicolumn{2}{c}{XO-2S} &   \colhead{}& \multicolumn{4}{c}{XO-2N}  \\
\cline{2-3} \cline{5-8} \\
\colhead{} & \colhead{this work} & \colhead{Burke et al.\,2007} & \colhead{} & \colhead{this work} & \colhead{Burke et al.\,2007} & \colhead{Ammler-von Eiff et al.\,2009} &\colhead{Torres et al.\,2012}}
\startdata
T$_{\rm{eff}}$ (K) & 5547$\pm$59& 5500$\pm$32 &    &5343$\pm$78 & 5340$\pm$32 & 5350$\pm$72 & 5450$\pm$75 \\
log $g$ (cgs) & 4.22 $\pm$0.24 & 4.62$\pm$0.05 &  & 4.49$\pm$0.25 & 4.48$\pm$0.05 & 4.14 $\pm$0.22 & 4.45 $\pm$0.02 \\
$\xi$ (km s$^{-1}$)& 1.24$\pm$0.07 & \nodata &  &1.22$\pm$0.09 & \nodata &1.10$\pm$0.08 & \nodata \\

[Fe/H] & 0.28$\pm$0.14 & 0.47$\pm$0.02 & & 0.39$\pm$0.14  &0.45$\pm$0.02 & 0.42 $\pm$0.07 & 0.27$\pm$0.11 \\

[C/H] & 0.26$\pm$0.11 & \nodata & & 0.42$\pm$0.12 & \nodata & \nodata & \nodata \\

[Ni/H] & 0.38$\pm$0.04 & 0.52$\pm$0.02 &  & 0.44$\pm$0.04 & 0.50$\pm$0.02 & \nodata & \nodata \\

[O/H] & 0.18$\pm$0.15 & \nodata & & 0.34$\pm$0.16 & \nodata & \nodata & \nodata \\

C/O & 0.60$\pm$0.19 & \nodata &   & 0.60$\pm$0.20 & \nodata & \nodata & \nodata \\

\enddata

\tablenotetext{a}{Adopted solar parameters: T$_{\rm{eff}}=$5777 K, log $g=$4.44, and $\xi=$1.38 km s$^{-1}$.}
\tablecomments{In this table the data listed as Torres et al.\,2012 is only that derived from their MOOG-style analysis.}
\end{deluxetable}

\clearpage


\end{document}